\documentclass[
prd
,showpacs ,amssymb,nobibnotes,aps,eqsecnum]{revtex4}
\usepackage{graphicx}
\input{epsf}

\usepackage{amsmath,amssymb}
\usepackage{bm}

\newcommand{\dalm}{\kern1pt\vbox{\hrule height 0.9pt\hbox{\vrule width 0.9pt
\hskip 2.5pt\vbox{\vskip 5.5pt}\hskip 3pt\vrule width 0.3pt}\hrule height 0.3pt}
\kern1pt}

\begin{document}



\title{Probing Tensor-Vector-Scalar Theory with Gravitational Wave Asteroseismology}

\author{Hajime Sotani} \email{sotani@astro.auth.gr}
\affiliation{
Theoretical Astrophysics, University of T\"{u}bingen, Auf der Morgenstelle 10, T\"{u}bingen 72076, Germany
}

\date{\today}

\begin{abstract}
In order to examine the gravitational waves emitted from the neutron stars in the tensor-vector-scalar
(TeVeS) theory, we derive the perturbation equations for relativistic stars, where for simplicity we omit
the perturbations of vector field. That is, we consider the perturbations of scalar and tensor fields.
With this assumption, we find that the axial gravitational waves, which are corresponding to the oscillations
of spacetime ($w$ modes), are independent from the perturbations of scalar field and the effects of scalar field
can be mounted only via the background properties. Using two different equations of state, we calculate
the complex eigenfrequencies of axial $w$ modes and find that the dependences of frequencies
on the stellar compactness are almost independent from 
the adopted equation of state and the parameter in TeVeS. Additionally, these dependences of frequencies of
axial $w$ modes in TeVeS is obviously different from those expected in the general relativity. Thus the direct
observations of gravitational waves could reveal the gravitational theory in the strong-field regime.
\end{abstract}

%
%
\maketitle
\section{Introduction}
\label{sec:I}

As an alternative gravitational theory, the tensor-vector-scalar (TeVeS) theory has attracted considerable
attention. This theory is proposed by Bekenstein \cite{Bekenstein2004} as a relativistic theory of Modified
Newtonian Dynamics (MOND) \cite{Milgrom1983}, and TeVeS reproduces MOND in the weak acceleration limit.
The most important advantage to adopt TeVeS might be the point to be possible to explain many galactic and
cosmological observations without the need for dark matter \cite{Note01}.
In TeVeS, one has successfully explained the galaxy rotation curves and
Tully-Fisher law without the existence of dark matter \cite{Bekenstein2004}.
Additionally, TeVeS is possible to explain not only the strong gravitational lensing \cite{Chen2006},
but also the galaxy distribution through an evolving Universe \cite{Dodelson2006} without cold dark matter.
On the other hand, in the strong-field regime of TeVeS, Giannios found the black hole solution by solving
the field equations for static, spherically symmetric spacetime in vacuum \cite{Giannios2005},
i.e., this corresponds to the Schwarzschild solution in the general relativity (GR), where he found two
distinct branches of solutions dependent on the form of the vector field. Subsequently, Sagi and Bekenstein
generalized the Schwarzschild solution in TeVeS to the Reissner-Nordstr\"om solution \cite{Sagi2008},
while Lasky, Sotani, and Giannios derived the Tolman-Oppenheimer-Volkoff (TOV) equations and
they produced the neutron star models in TeVeS \cite{Paul2008}.

Recently, it was discussed the possibilities of distinguishing TeVeS from GR.
For example, it was suggested that one can distinguish the gravitational theory with the redshift
of the atomic spectral lines emanating from the surface of neutron stars \cite{Paul2008} or
with the observations of neutron star oscillations via the emitted gravitational waves \cite{Sotani2009a}.
Additionally, Desai, Kahya, and Woodard pointed out another test of TeVeS \cite{Desai2008},
where they showed that there is an appreciable difference in the Shapiro delays
of gravitational waves and photons or neutrinos from the same source and suggested
the possibility to test of TeVeS by observing this difference emitted from gamma
ray bursts or core-collapse supernovae. In this article, we examine whether observations of
gravitational waves emitted from the neutron stars can provide an alternative way of probing
the gravitational theory in the strong-field regime, where we focus especially on the gravitational waves
associated with the oscillation of spacetime itself.

In fact, the tests of gravitational theories in the strong-field regime are quite important, because
the gravitational theories in the strong-field regime are still largely unconstrained by observations
in contrast to those in the weak-field regime, which have been subject to numerous experimental tests.
However, owing to the developments in observational technology, it is becoming possible to observe
compact objects with high accuracy. Via the observations of X-rays and/or gamma rays emitted from
compact objects, one can know the properties of compact objects and could use as a direct test of
the gravitational theory in the strong-field regime. As an alternative way to observe compact objects,
the gravitational waves are also expected to obtain the raw information of the compact objects.
With these observational properties, it might be possible to distinguish the gravitational theory
in the strong-field regime \cite{Psaltis2008,DeDeo2003,Sotani2004}.

Meanwhile,
it has been suggested that the observations of gravitational waves can provide a unique tool not only
for estimating the stellar parameters such as mass, radius, rotation rate, magnetic field, and equation
of state (e.g., \cite{Andersson1998,Sotani2001,Sotani2003,GPM2004,Erich2008,AGH2008}),
but also for verifying the gravitational theory, which is called ``gravitational-wave asteroseismology".
Furthermore, the detailed analysis of the gravitational waves also makes it possible to determine the radius
of accretion disk around supermassive black hole \cite{Sotani2006} or to know the magnetic effects during
the stellar collapse \cite{SYK2007}. The eigenmodes, which are mainly excited during the formation of a
neutron star or during the starquakes and emit detectable gravitational waves with the ground-based
gravitational wave detectors such as LIGO, GEO600, VIRGO, and TAMA300, are the fluid $f$ and $p$ modes
and the $w$ modes \cite{Kokkotas1992}, which are associated with oscillations of the spacetime \cite{Kokkotas2001}.
The possibility to distinguish TeVeS from GR by using the fluid oscillations ($f$ and $p$ modes)
has been already discussed in \cite{Sotani2009a}, where the Cowling approximation was adopted.
Thus in this article we focus on the $w$ modes. The $w$ modes are similar to quasinormal modes of black holes.
They have higher frequencies and shorter damping times than the fluid modes, i.e., in GR the typical frequencies
are around 7 -- 12 kHz and damping times are order of 0.1 ms. In general, the oscillations on the spherically
symmetric spacetime can be classified as axial and polar depending on their parity. Since the axial $w$ modes
are known to have the same qualitative behavior as the polar $w$ modes in GR, in this article we will examine
only the axial $w$ modes. Moreover, for simplicity, we consider the perturbations of scalar and tensor
fields while those of vector field will be omitted in this article. The more complicated analysis
of the polar $w$ modes and/or including the perturbations of vector field will be seen somewhere.

This article is organized as follows. In the next section, we describe our notation and
briefly introduce the theoretical framework of TeVeS.
In Sec. \ref{sec:III} we derive the perturbation equations for the axial perturbations.
Then the oscillation spectra of neutron stars in TeVeS are shown
in Sec. \ref{sec:IV}, finally we discuss the results related to gravitational wave asteroseismology
in Sec. \ref{sec:V}. In this article, we adopt the unit of $c=G=1$, where $c$
and $G$ denote the speed of light and the gravitational constant, respectively, and
the metric signature is $(-,+,+,+)$.

\section{Stellar Models in TeVeS}
\label{sec:II}

In this section, we mention the stellar models in TeVeS.
TeVeS is based on three dynamical gravitational fields, such as an Einstein metric $g_{\mu\nu}$, 
a timelike 4-vector field ${\cal U}^\mu$, and a scalar field $\varphi$,
in addition to a nondynamical scalar field $\sigma$.
The vector field fulfills the normalization condition
$g_{\mu\nu}{\cal U}^\mu{\cal U}^\nu=-1$ and the physical metric is given by
\begin{gather}
 \tilde{g}_{\mu\nu} = e^{-2\varphi}\left(g_{\mu\nu} + {\cal U}_\mu{\cal U}_\nu\right)
     - e^{2\varphi}{\cal U}_\mu{\cal U}_\nu. \label{gg}
\end{gather}
All quantities in the physical frame are denoted with a tilde, and any
quantity without a tilde is in the Einstein frame.
Varying the total action with respect to $g^{\mu\nu}$,
one can obtain the field equations for the tensor field
\begin{equation}
 G_{\mu\nu} = 8\pi G \left[\tilde{T}_{\mu\nu}+\left(1-e^{-4\varphi}\right){\cal U}^\alpha
               \tilde{T}_{\alpha(\mu}{\cal U}_{\nu)}+\tau_{\mu\nu}\right]+\Theta_{\mu\nu},
               \label{Einstein}
\end{equation}
where $\tilde{T}_{\mu\nu}$ is the energy-momentum
tensor in the physical frame, $\tilde{T}_{\alpha(\mu}{\cal U}_{\nu)}\equiv\tilde{T}_{\alpha\mu}{\cal U}_{\nu}
+\tilde{T}_{\alpha\nu}{\cal U}_{\mu}$, and $G_{\mu\nu}$ is the Einstein tensor
in the Einstein frame, while the other sources $\tau_{\mu\nu}$ and $\Theta_{\mu\nu}$
are given by Eqs. (2.4) and (2.5) in \cite{Sotani2009a}.
Notice that the conservation of energy-momentum is given in
the physical frame as $\tilde{\nabla}_\mu \tilde{T}^{\mu\nu}=0$.
%
%
%
%
%
Additionally, by varying the total action with respect to ${\cal U}_\mu$ and $\varphi$,
one obtains the field equations for the vector and scalar fields as Eqs. (2.6) and (2.7)
in \cite{Sotani2009a}, which include two positive dimensionless parameters, $k$ and ${\cal K}$.
These are the coupling parameters for the scalar and vector fields respectively.

A static, spherically symmetric metric in Einstein frame can be expressed as
\begin{equation}
 ds^2=g_{\alpha\beta}dx^\alpha dx^\beta = -e^{\nu(r)}dt^2 + e^{\zeta(r)}dr^2 + r^2 d\Omega^2,
\end{equation}
where $d\Omega^2 = d\theta^2+\sin^2\theta d\phi^2$ and $e^{-\zeta}=1-2m(r)/r$,
while the vector field can be given as ${\cal U}^\mu=\left({\cal U}^t(r),{\cal U}^r(r),0,0\right)$.
In this article we set ${\cal U}^r$=0, because it has shown that in vacuum, the parameterized
post-Newtonian (PPN) coefficients for a spherically symmetric, static spacetime
with a non-zero ${\cal U}^{r}$ can violate observational restrictions \cite{Giannios2005}.
Then with the normalization condition one can show that ${\cal U}^\mu=\left(e^{-\nu/2},0,0,0\right)$.
With this vector field, the physical metric is
\begin{equation}
 d\tilde{s}^2=\tilde{g}_{\alpha\beta}dx^\alpha dx^\beta
     = -e^{\nu+2\varphi}dt^2 + e^{\zeta-2\varphi}dr^2 + e^{-2\varphi}r^2 d\Omega^2.
\end{equation}
Finally we assume that the stellar matter consists of a perfect fluid
\begin{equation}
 \tilde{T}_{\mu\nu} = \left(\tilde{\rho}+\tilde{P}\right)\tilde{u}_\mu\tilde{u}_\nu
     + \tilde{P}\tilde{g}_{\mu\nu},
\end{equation}
where $\tilde{u}_\mu$, $\tilde{\rho}$, and $\tilde{P}$ are the four-velocity of the fluid,
the total energy density, and the pressure in the physical frame.
The concrete neutron star models in TeVeS can be seen in the previous papers \cite{Paul2008,Sotani2009a}.

To construct the stellar models in TeVeS, three new parameters, $k$, ${\cal K}$ and $\varphi_c$,
are introduced with respect to GR, where $\varphi_c$ denotes the cosmological value of the scalar field.
The value of $k$ has a tightly constraint as $k\sim 0.03$ by both cosmological models and also planetary
motions in the solar system, while the cosmological considerations imply that
the value of $\varphi_c$ is restricted to $0\le\varphi_c\ll 1$ \cite{Bekenstein2004}.
With respect to the value of $\varphi_c$, it was shown that $\varphi_c$ could have a minimum value of around
$0.001$, which is based on the causality issues inside the star \cite{Paul2008}.
Anyway, since it was also found in \cite{Paul2008} that the neutron star models are almost independent from
the values of $k$ and $\varphi_c$, in this article we make examinations with $k=0.03$ and $\varphi_c=0.003$.
On the other hand, although we pointed out in \cite{Sotani2009a} the possibility to determine the value
of ${\cal K}$ with the observations of gravitational waves related to the stellar oscillations, so far
there is no severe restriction on ${\cal K}$ except that ${\cal K}$ should be in the range of $0<{\cal K}<2$
\cite{Sagi2008,Paul2008}. Thus in this article we adopt various values of ${\cal K}$ in
the range of $0<{\cal K}<2$ and study the dependence of gravitational waves on the parameter ${\cal K}$.
Furthermore, as equilibrium stellar models, in this article we adopt the similar models in \cite{Sotani2009a},
i.e., two different EOSs such as EOS A and EOS II.

At the end of this section, we introduce two useful properties, i.e., the total Arnowitt-Deser-Misner (ADM)
mass $M_{\rm ADM}$ and the scalar mass $M_{\varphi}$, which are defined as
\begin{gather}
 M_{\rm ADM} = \left(m_c + \frac{kGM_\varphi}{4\pi}\right)e^{-\varphi_c}, \\
 M_\varphi = 4\pi\int_0^rr^2\left(\tilde{\rho}+3\tilde{P}\right)e^{(\nu+\zeta)/2-2\varphi}dr,
\end{gather}
where $m_c$ is the mass function evaluated at radial infinity.
Notice that the scalar mass is constant outside the star, while with ADM mass one can describe the 
asymptotic behavior of the physical metric as
\begin{gather}
 \tilde{g}_{tt} = -1 + \frac{2M_{\rm ADM}}{\tilde{r}} + {\cal O}\left(\frac{1}{\tilde{r}^2}\right), \\
 \tilde{g}_{rr} =  1 + \frac{2M_{\rm ADM}}{\tilde{r}} + {\cal O}\left(\frac{1}{\tilde{r}^2}\right),
\end{gather}
where $\tilde{r}\equiv re^{-\varphi_c}$.

\section{Perturbation equations}
\label{sec:III}

As mentioned before, for simplicity we assume that $\delta{\cal U}_i=0$ in this article,
where the subscript $i$ corresponds to $i=r$, $\theta$, and $\phi$.
In fact there is no physical reason for this assumption. But in order to construct the neutron star models with the observed masses, the value of the coupling constant for vector field, ${\cal K}$, should be smaller than at most ${\cal K}\simeq0.8$, which depends on the adopted EOS \cite{Paul2008,Sotani2009a}. So since one could expect that with smaller value of ${\cal K}$ the effect of the vector-field perturbation on the spacetime oscillations might be small, in this paper as a first step we neglect this type of perturbation to make our problem simple. (Also see in Appendix \ref{sec:appendix_0} for the effect of the vector-field perturbation on the spacetime oscillations.)
It is noteworthy that $\delta {\cal U}^\mu\ne 0$ and $\delta{\cal U}_t\ne0$ even with the assumption
that $\delta{\cal U}_i=0$, which are results from the metric perturbations. Under this situation,
the perturbations of the scalar and vector fields are expressed as
\begin{align}
 \delta \varphi &= \delta \varphi(t,r) Y_{lm}, \\
 \delta {\cal U}_\mu &= \left(\frac{1}{2}e^{-\nu/2}\delta g_{tt},0,0,0\right),
\end{align}
while, using the Regge-Wheeler gauge, the perturbed metric tensor
in the physical frame is given as
\begin{equation}
 \delta\tilde{g}_{\mu\nu}=\tilde{h}_{\mu\nu}^{(-)}+\tilde{h}_{\mu\nu}^{(+)},
\end{equation}
where $\tilde{h}_{\mu\nu}^{(-)}$ denotes the {\em axial} part of metric perturbations
\begin{eqnarray}
\tilde{h}_{\mu\nu}^{(-)}&=&\sum_{l=2}^{\infty} \sum_{m=-l}^{l}\left(
 \begin{array}{cccc}
 0  &  0 & -h_{0,lm} {\sin^{-1}\theta} \partial_{\phi} & h_{0,lm} \sin\theta \, \partial_{\theta} \\
 0  &  0 & -h_{1,lm} {\sin^{-1}\theta} \partial_{\phi} & h_{1,lm} \sin\theta \, \partial_{\theta} \\
 \ast & \ast & 0 & 0 \\
 \ast & \ast & 0 & 0 \\
 \end{array}
 \right) Y_{lm},
\end{eqnarray}
and $\tilde{h}_{\mu\nu}^{(+)}$ denotes the {\em polar} part of metric perturbations
\begin{eqnarray}
\tilde{h}_{\mu\nu}^{(+)}&=&\sum_{l=2}^{\infty} \sum_{m=-l}^{l}\left(
 \begin{array}{cccc}
 H_{0,lm} e^{\nu}  &  H_{1,lm}           & 0          & 0 \\
 *                 &  H_{2,lm} e^{\zeta} & 0          & 0 \\
 0                 &  0                  & r^2 K_{lm} & 0 \\
 0                 &  0                  & 0          & r^2 K_{lm} \sin^2 \theta \\
 \end{array}
 \right) Y_{lm} \, .
\end{eqnarray}
Here, the functions $h_{0,lm}$, $h_{1,lm}$, $H_{0,lm}$, $H_{1,lm}$,
$H_{2,lm}$, and $K_{lm}$ describing the spacetime perturbations have
only radial and temporal dependence while
$Y_{lm}=Y_{lm}(\theta,\phi)$ is the spherical harmonic function.
Since, from Eq. (\ref{gg}), the perturbed metric tensor in the physical frame
is connected to that in the Einstein frame $\delta g_{\mu\nu}$ as
\begin{equation}
 \delta \tilde{g}_{\mu\nu} = -2 e^{-2\varphi}\left(g_{\mu\nu} + {\cal U}_\mu{\cal U}_\nu\right)\delta\varphi
    + e^{-2\varphi}\left(\delta g_{\mu\nu} + \delta {\cal U}_\mu{\cal U}_\nu
    + {\cal U}_\mu\delta {\cal U}_\nu\right) - 2 e^{2\varphi}{\cal U}_\mu{\cal U}_\nu\delta\varphi
    - e^{2\varphi}\left(\delta {\cal U}_\mu{\cal U}_\nu + {\cal U}_\mu\delta{\cal U}_\nu\right),
    \label{p_Einstein}
\end{equation}
the perturbed metric tensor in the Einstein frame is
\begin{align}
\delta g_{\mu\nu} = \sum_{l=2}^{\infty} \sum_{m=-l}^{l}\left(
 \begin{array}{cccc}
 \left(e^{-2\varphi}H_{0,lm} + 2\delta\varphi \right)e^{\nu} &  e^{2\varphi}H_{1,lm}
 & -e^{2\varphi}h_{0,lm} \sin^{-1} \theta \partial_{\phi}
 & e^{2\varphi}h_{0,lm} \sin\theta \, \partial_{\theta} \\
 \ast & \left(e^{2\varphi}H_{2,lm} + 2\delta\varphi \right) e^{\zeta}
 & -e^{2\varphi}h_{1,lm} \sin^{-1} \theta \partial_{\phi}
 & e^{2\varphi}h_{1,lm} \sin\theta \, \partial_{\theta} \\
 \ast & \ast & \left(e^{2\varphi}K_{lm} + 2\delta\varphi\right) r^2 & 0 \\
 \ast & \ast & 0 & \left(e^{2\varphi}K_{lm} + 2\delta\varphi\right) r^2 \sin^2\theta \\
 \end{array}
 \right) Y_{lm}.
\label{Eq:h_Einstein}
\end{align}
This expression of $\delta g_{\mu\nu}$ is used
when the perturbations in the Einstein frame will be transformed back to the physical frame.
Now, by defining the new set of perturbation functions, $\hat{H}_{0,lm}$,
$\hat{H}_{1,lm}$, $\hat{H}_{2,lm}$, $\hat{K}_{lm}$, $\hat{h}_{0,lm}$, and $\hat{h}_{1,lm}$, as
follows
\begin{align}
 \hat{H}_{0,lm} &= e^{-2\varphi}H_{0,lm} + 2\delta\varphi, \\
 \hat{H}_{1,lm} &= e^{2\varphi}H_{1,lm}, \\
 \hat{H}_{2,lm} &= e^{2\varphi}H_{2,lm} + 2\delta\varphi, \\
 \hat{K}_{lm}   &= e^{2\varphi}K_{lm} + 2\delta\varphi, \\
 \hat{h}_{0,lm} &= e^{2\varphi}h_{0,lm}, \\
 \hat{h}_{1,lm} &= e^{2\varphi}h_{1,lm},
\end{align}
the perturbed metric in the Einstein frame is simplified considerably and reduced
to the ``standard" Regge-Wheeler form of perturbed spherical metric.
Here we emphasize that the scalar perturbation $\delta\varphi$ is linked only with
the polar perturbations $H_{0,lm}$, $H_{1,lm}$, $H_{2,lm}$, and $K_{lm}$, while
the axial perturbations $h_{0,lm}$ and $h_{1,lm}$ are affected only by the contribution
of the scalar field to the background.

The perturbation equations for the gravitational waves
can be obtained by taking the variation of tensor field
equations (\ref{Einstein}). In order to derive the perturbation equations, we define
the variations of pressure and energy density in physical frame  as
\begin{align}
 \delta\tilde{P} &= \delta\tilde{P}Y_{lm}, \\
 \delta\tilde{\rho} &= \delta\tilde{\rho}Y_{lm},
\end{align}
while the variation of four-velocity in physical frame as
\begin{align}
 \delta\tilde{u}^t &= \frac{1}{2}e^{-3\varphi-\nu/2}H_0Y_{lm}, \\
 \delta\tilde{u}^r &= \frac{1}{r^2}e^{-\varphi-\nu/2}WY_{lm}, \\
 \delta\tilde{u}^\theta &= \frac{1}{r^2}e^{-\varphi-\nu/2}\left(V\partial_\theta Y_{lm}
     - u\frac{1}{\sin\theta}\partial_\phi Y_{lm}\right), \\
 \delta\tilde{u}^\phi   &= \frac{1}{r^2\sin^2\theta}e^{-\varphi-\nu/2}\left(V\partial_\phi Y_{lm}
     + u\sin\theta\partial_\theta Y_{lm}\right).
\end{align}
where $\delta\tilde{P}$, $\delta\tilde{\rho}$, $W$, $V$, and $u$ are functions of $t$ and $r$.
Using these definitions, from the $t\theta$, $t\phi$, $r\theta$, and $r\phi$ components of linearized
Einstein equations, we get
\begin{gather}
 \sum_{l,m}\left\{\alpha_{lm}^{(J)}\partial_\theta Y_{lm}
     + \beta_{lm}^{(J)}\frac{1}{\sin\theta}\partial_\phi Y_{lm}\right\} = 0 \ \ (J=0,1),
       \label{t-theta} \\
 \sum_{l,m}\left\{\beta_{lm}^{(J)}\partial_\theta Y_{lm}
     - \alpha_{lm}^{(J)}\frac{1}{\sin\theta}\partial_\phi Y_{lm}\right\} = 0 \ \ (J=0,1),
\end{gather}
where the explicit expressions of $\alpha_{lm}^{(J)}$ and $\beta_{lm}^{(J)}$ are given in Appendix
\ref{sec:appendix_1}. Notice that $\alpha_{lm}^{(J)}$ are some linear combinations of polar perturbation
functions, while $\beta_{lm}^{(J)}$ consist of only axial perturbation functions.
Furthermore, from the $\theta\phi$ component and the subtraction of $\theta\theta$ and $\phi\phi$
components, one obtains two more equations
\begin{gather}
 \sum_{l,m}\left\{s_{lm}X_{lm} - t_{lm}\sin\theta W_{lm}\right\} = 0, \\
 \sum_{l,m}\left\{t_{lm}X_{lm} + s_{lm}\sin\theta W_{lm}\right\} = 0, \label{phiphi}
\end{gather}
where $s_{lm}$ and $t_{lm}$ describe polar and axial type perturbations, respectively
(see Appendix \ref{sec:appendix_1}), while $X_{lm}$ and $W_{lm}$ are defined as
\begin{align}
 X_{lm} &= 2\partial_\phi\left(\partial_\theta - \frac{\cos\theta}{\sin\theta}\right)Y_{lm}, \\
 W_{lm} &= \left(\partial_\theta^2 - \frac{\cos\theta}{\sin\theta}\partial_\theta
     - \frac{1}{\sin^2\theta}\partial_\phi^2\right)Y_{lm}.
\end{align}
Additionally, from the $tt$, $tr$, $rr$ components and the sum of the $\theta\theta$
and $\phi\phi$ components, one can get four more equations
\begin{equation}
 \sum_{l,m}A_{lm}^{(I)}Y_{lm}=0\ \ (I=0,1,2,3), \label{tt}
\end{equation}
but we do not care these expressions in this article
since these equations are corresponding to polar perturbations.

Then, by taking the product of Eqs. (\ref{t-theta}) -- (\ref{phiphi}) and (\ref{tt}) with $\bar{Y}_{lm}$,
integrating over the solid angle, and paying attention to the fixed values of $l$ and $m$,
we get ten partial differential equations in the variables $t$ and $r$ as
\begin{gather}
 A_{lm}^{(I)}=0,\ \  \alpha_{lm}^{(J)}=0,\ \  s_{lm}=0, \label{polar} \\
 \beta_{lm}^{(J)}=0,\ \  t_{lm}=0, \label{axial}
\end{gather}
where $I=0,1,2,3$ and $J=0,1$.
Here Eqs. (\ref{polar}) describe the polar perturbations, while Eqs. (\ref{axial})
describe the axial perturbations. It is worth noticing that the analytic expressions
for Eqs. (\ref{axial}), i.e., Eqs. (\ref{beta0}), (\ref{beta1}) and (\ref{tlm}),
do not involve the perturbation of scalar field $\delta \varphi$. Thus the scalar
perturbation is coupled only to the gravitational waves with polar parity.

Combining Eqs. (\ref{axial}), one can easily derive a wave equation for the axial perturbations as
\begin{equation}
 \ddot{\Phi} - e^{(\nu-\zeta)/2}\left(e^{(\nu-\zeta)/2}\Phi'\right)'
     + e^{\nu-\zeta}\left[\nu'' + \frac{\nu'}{2}\left(\nu'-\zeta'+\frac{5}{r}\right)
     - \frac{1}{r^3}e^{\zeta}\left\{rm' + 7m - l(l+1)r + 32\pi G\tilde{P}r^3e^{-2\varphi}\right\}\right]
       \Phi = 0, \label{eq:wave}
\end{equation}
where the new function $\Phi(t,r)$ is introduced, which is defined as
$\hat{h}_1 \equiv e^{(\zeta-\nu)/2}r\Phi$.
Note that if $k={\cal K}=0$ this equation reduces to the standard wave equation describing axial
perturbations in GR \cite{CF1991}.
As mentioned the above, the wave equation (\ref{eq:wave}) does not include
the perturbations of scalar field and the effects of scalar field will enter only via the background
properties. That is, the axial gravitational waves can be studied independently neither from polar
gravitational waves nor from the scalar field perturbations. This axial gravitational waves are well-known
as $w$ modes \cite{Kokkotas1992}, which are quasinormal modes describing the pure spacetime oscillations
and similar to the quasinormal modes of black holes.

\section{Spacetime Perturbations in TeVeS}
\label{sec:IV}

In order to determine the quasinormal frequencies of the axial $w$ modes, one can think up two different
techniques. The first approach is the direct time evolutions of Eq. (\ref{eq:wave}) and the frequencies
will be determined by computing the Fourier transform of the signal at infinity,
while in the second approach, one assumes a harmonic time dependence of the perturbations and
the frequencies will be obtained by solving the eigenvalue problem with the appropriate boundary
conditions. The first approach is quite simple, but with this approach one can identify only those of the quasinormal
modes which are excited significantly, and the outcome depends strongly on the choice of the initial data. Additionally, using
time evolutions it is quite difficult to identify quasinormal modes that damp out very fast.
On the other hand, although the second approach is more complicated than the first one,
it is possible to calculate both slowly and strongly damped quasinormal modes. Thus,
in this article we adopt the second approach to determine the quasinormal frequencies of axial $w$ modes.
Assuming a harmonic time dependence as $\Phi(t,r)=\Phi(r)e^{i\omega t}$, the wave equation (\ref{eq:wave})
can be rewritten as
\begin{equation}
 \Phi'' + \frac{\nu'-\zeta'}{2}\Phi'
     + \left[\omega^2e^{\zeta-\nu} - \nu'' - \frac{\nu'}{2}\left(\nu'-\zeta'+\frac{5}{r}\right)
     + \frac{1}{r^3}e^{\zeta}\left\{rm' + 7m - l(l+1)r + 32\pi G\tilde{P}r^3e^{-2\varphi}\right\}\right]
       \Phi = 0. \label{eq:wave2}
\end{equation}
Then, with appropriate boundary conditions, the problem to solve becomes an eigenvalue one with respect
to the complex quasinormal frequencies $\omega$. Physically, the real and imaginary parts of the complex
frequency are corresponding to the oscillation frequency and the damping rate of each eigenmode,
respectively. The imposed boundary conditions are that $\Phi$ should
be regular at the stellar center and that there are no incoming waves at infinity. In fact, near the stellar
center, one can show that $\Phi$ has a behavior of the form as
$\Phi(r) = \Phi_0 r^{l+1}\left(1+{\cal O}(r^2)\right)$, where $\Phi_0$ is some arbitrary constant.
With this boundary condition, we can just integrate the above differential equation inside the star.
On the other hand, outside the star, we use appropriate asymptotic expansions to meet the boundary
condition at infinity, where we adopt the Leaver's continued faction method \cite{Leaver1985}.
The concrete numerical procedure is described in Appendix \ref{sec:appendix_2}.

In Fig. \ref{fig:wmodes}, as examples, we show the eigenfrequencies of $w$ and $w_{\rm II}$ modes
for neutron star models with $M_{\rm ADM}=1.4M_\odot$, where the open symbols are corresponding to
the $w_{\rm II}$ modes and the solid ones denote the $w$ modes. From the observational point of view,
the lowest $w$ modes might to be relevant for the gravitational wave detectors \cite{Kokkotas1992},
while the higher $w$ modes, i.e., $w_2$, $w_3$, $w_4$, $\cdots$, are probably difficult to detect
if not impossible. As regarding to the $w_{\rm II}$ mode, the oscillation frequency is also suitable
for detection by ground-based interferometers, although they damp out very fast.
%
%
%
\begin{figure}[htbp]
\begin{center}
\begin{tabular}{cc}
\includegraphics[scale=0.45]{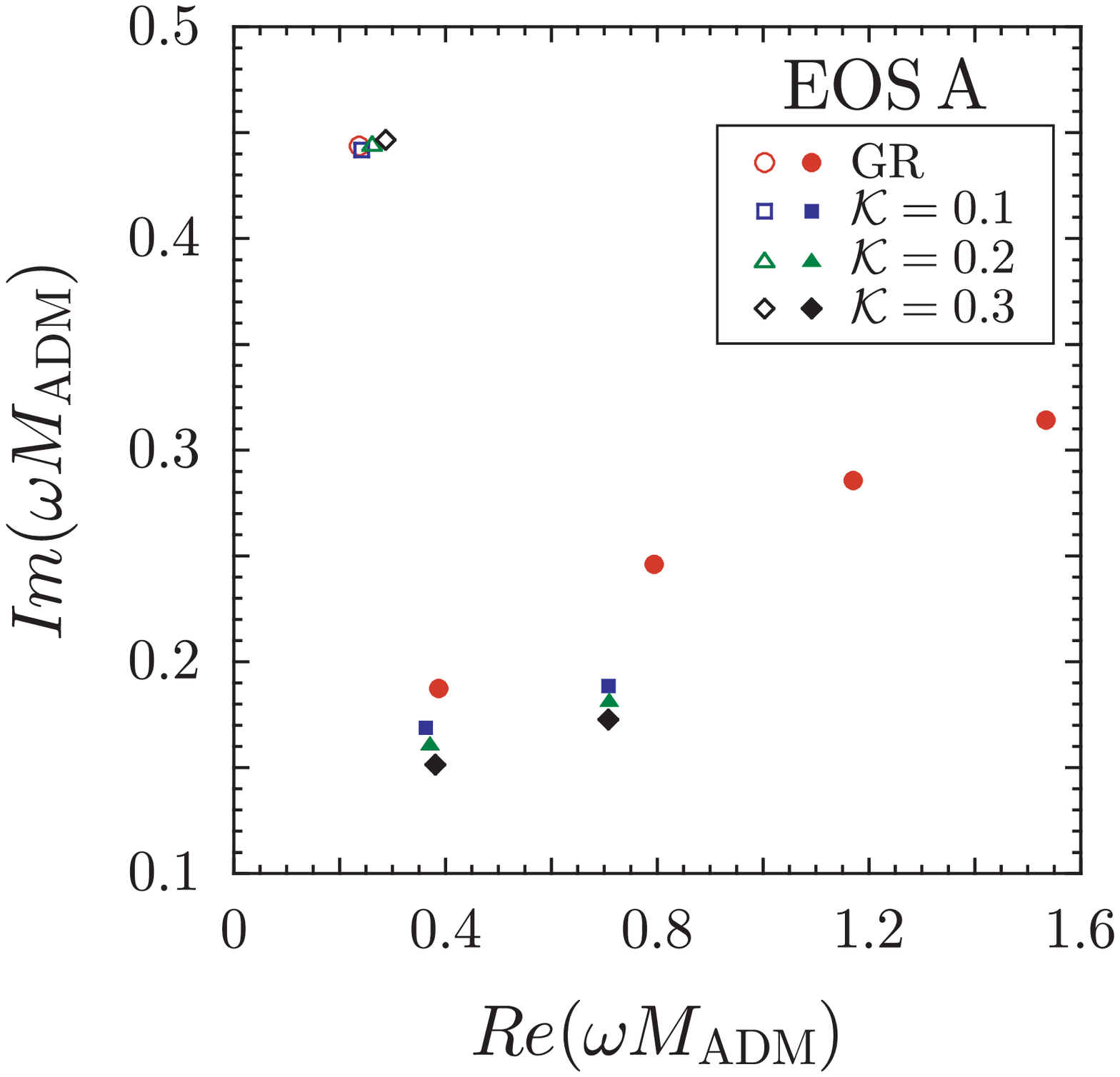} &
\includegraphics[scale=0.45]{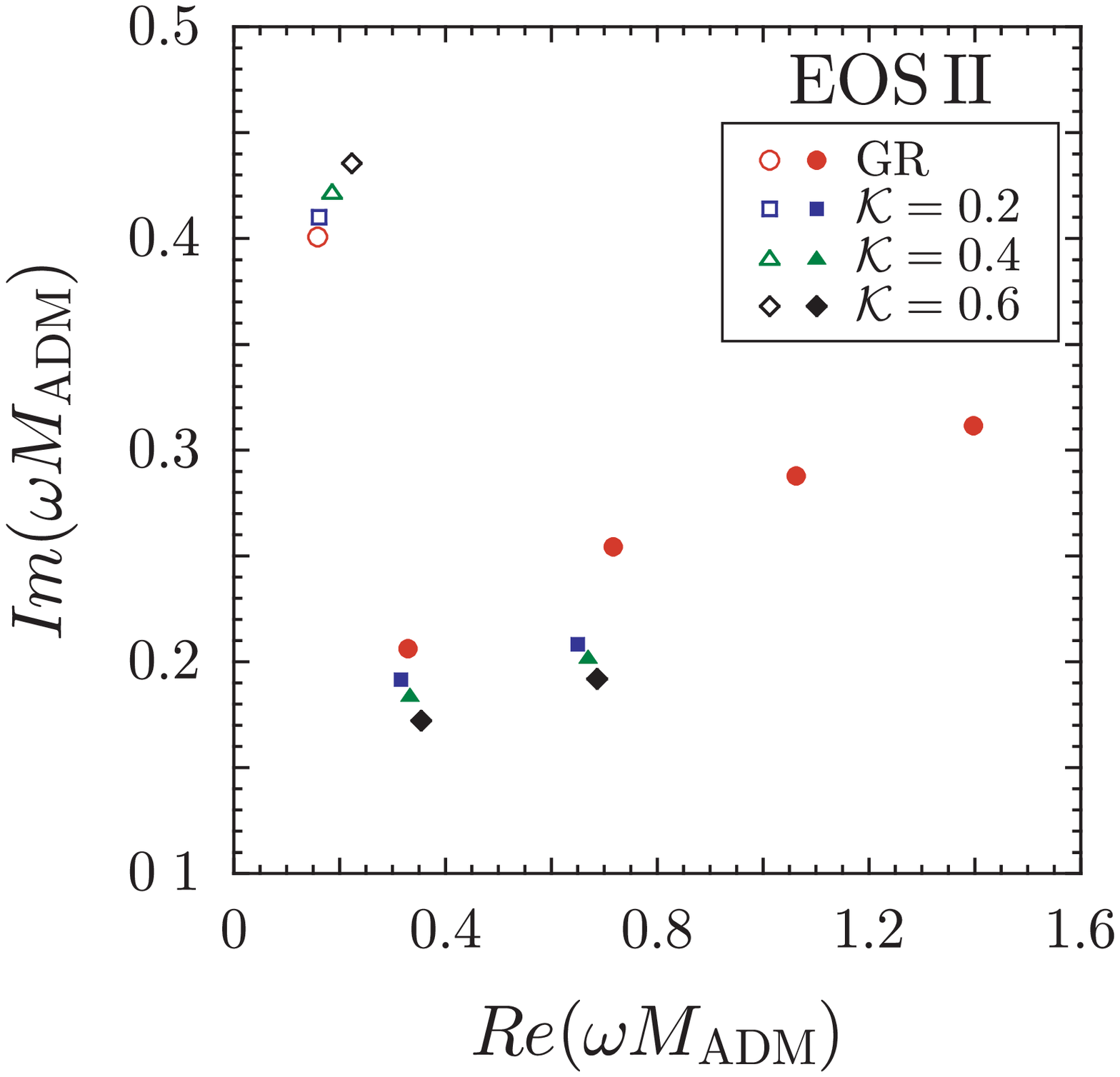} \\
\end{tabular}
\end{center}
\caption{
The complex frequencies of the axial $w$ and $w_{\rm II}$ modes for $l=2$, where the stellar masses are fixed
to be $M_{\rm ADM}=1.4M_\odot$. The left panel corresponds to EOS A and the right panel to EOS II.
In the figures, $w_{\rm II}$ modes are shown with the open symbols and $w$ modes with the solid ones.
The circles denote the frequencies in GR, while the other symbols denote those in TeVeS with different
values of ${\cal K}$.
}
\label{fig:wmodes}
\end{figure}

\begin{figure}[htbp]
\begin{center}
\begin{tabular}{cc}
\includegraphics[scale=0.45]{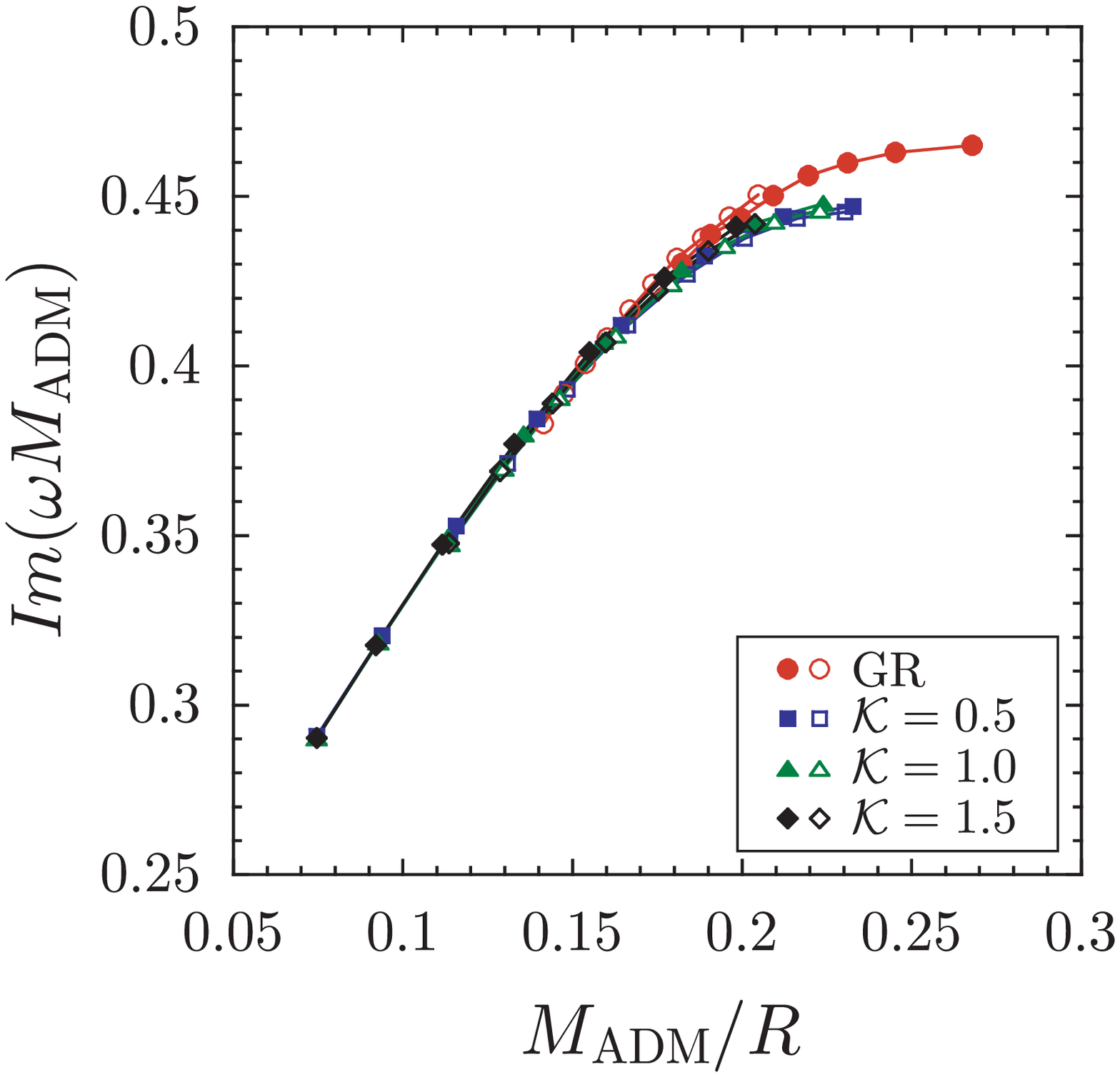} &
\includegraphics[scale=0.45]{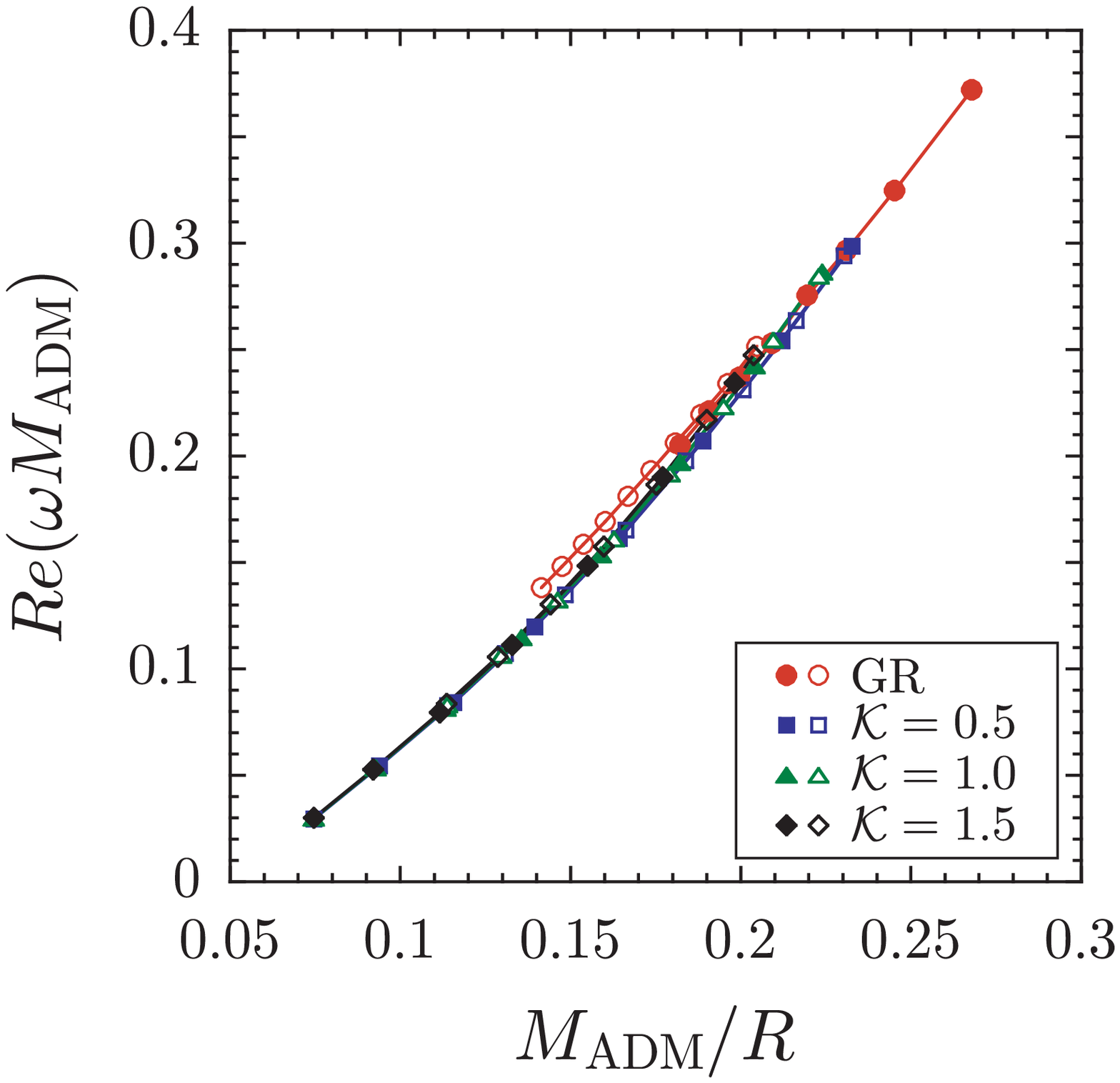} \\
\end{tabular}
\end{center}
\caption{
For $w_{\rm II}$ modes, the damping rates (left panel) and the oscillation frequencies (right panel)
as functions of stellar compactness $M_{\rm ADM}/R$.
The solid symbols correspond to the results for EOS A and the open ones to those for EOS II.
}
\label{fig:wII}
\end{figure}

\begin{figure}[htbp]
\begin{center}
\begin{tabular}{cc}
\includegraphics[scale=0.45]{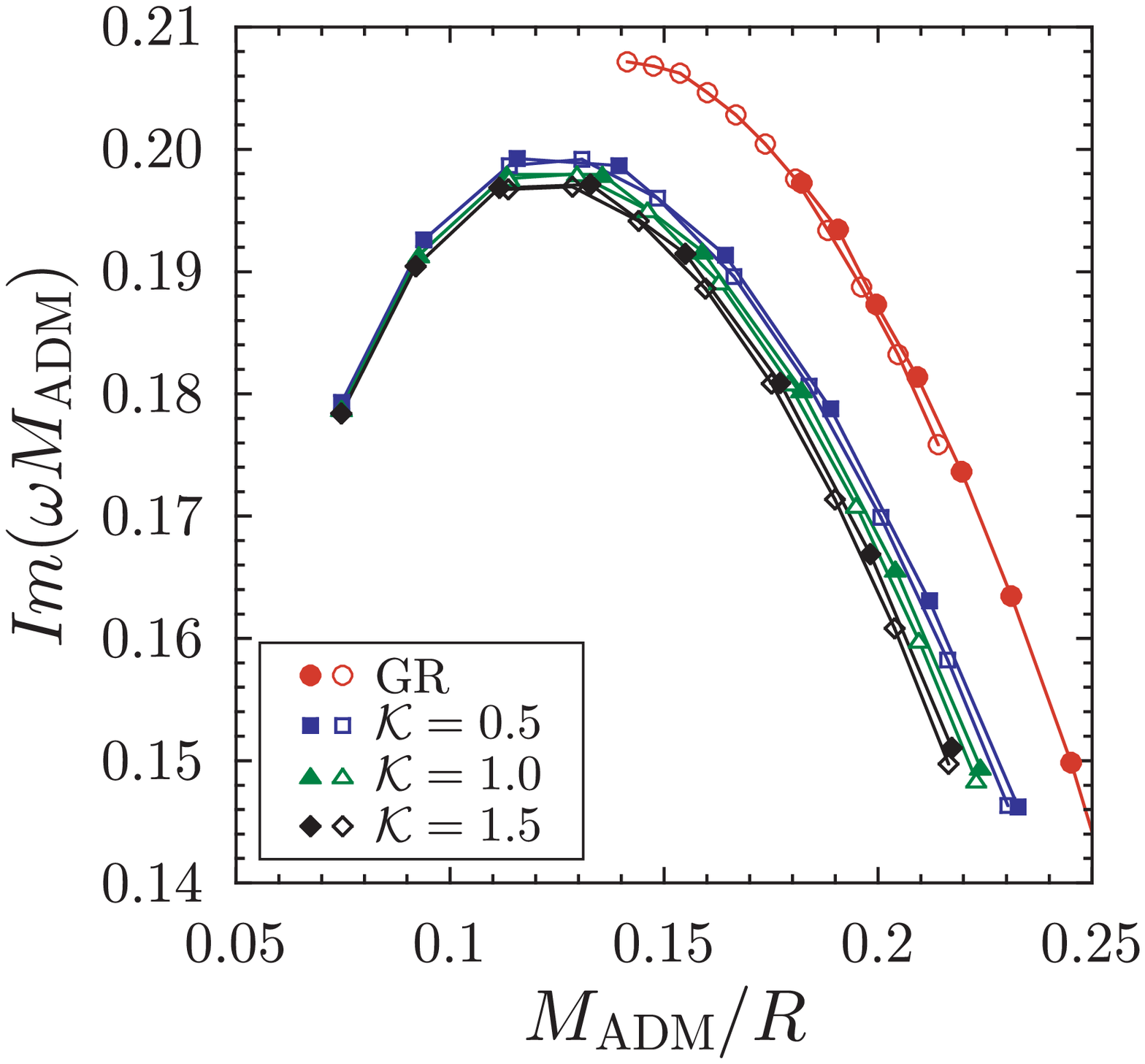} &
\includegraphics[scale=0.45]{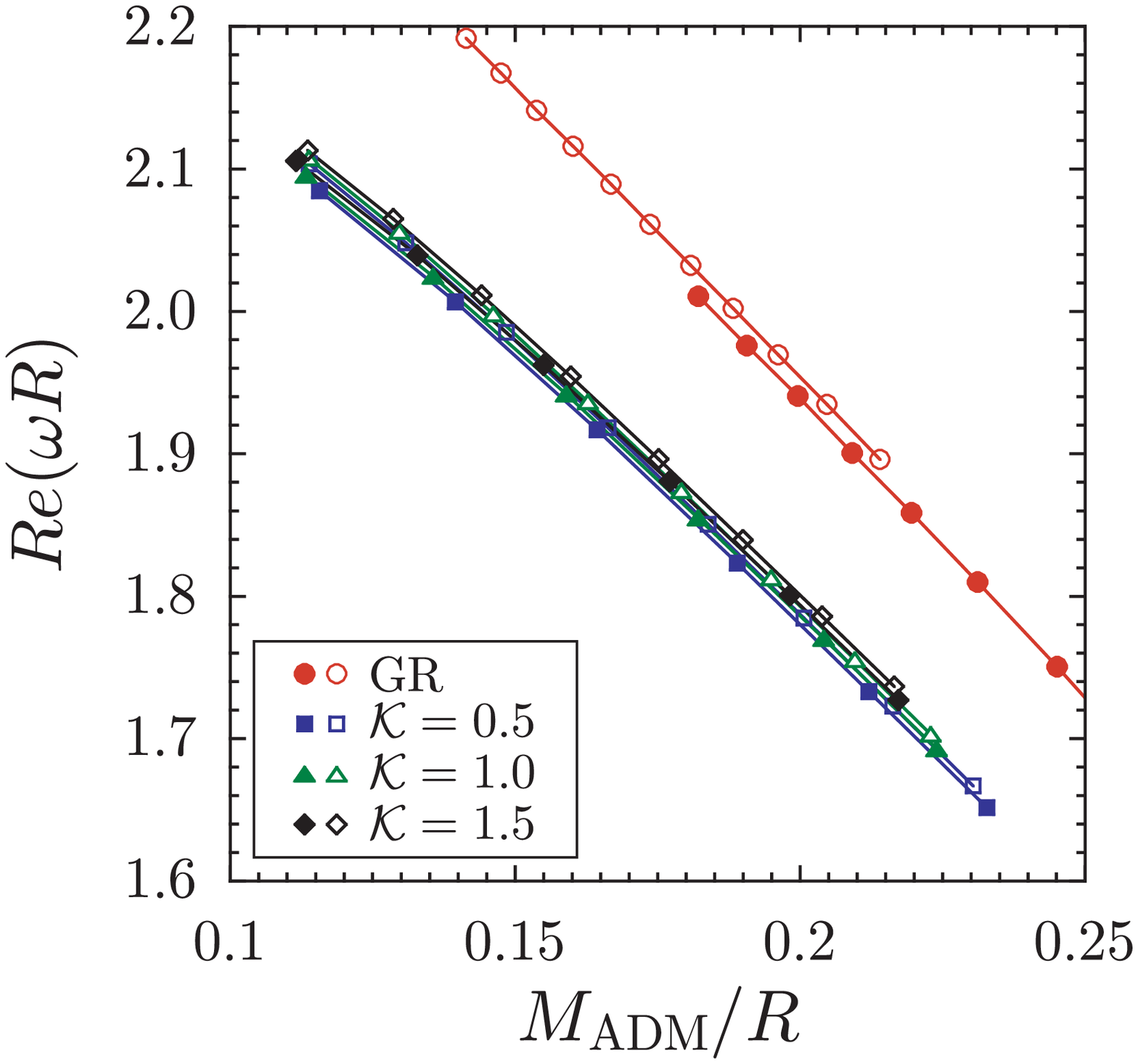} \\
\end{tabular}
\end{center}
\caption{
For $w_1$ modes, the damping rates (left panel) and the oscillation frequencies (right panel)
as functions of stellar compactness $M_{\rm ADM}/R$.
The solid symbols correspond to the results for EOS A and the open ones to those for EOS II.
}
\label{fig:w1}
\end{figure}

In the previous studies in GR, it is well known that the frequencies of $w$ and $w_{\rm II}$ modes
can be described as a function of stellar compactness $M_{\rm ADM}/R$ \cite{Kokkotas1992,Andersson1998}.
Similarly, we will examine the dependence of frequencies of $w_{\rm II}$ and $w$ modes on the stellar
compactness. Figs. \ref{fig:wII} and \ref{fig:w1} show the frequencies of axial $w_{\rm II}$ and $w_1$
as functions of the total stellar compactness, where the solid symbols are results with EOS A and the open
ones are those with EOS II. From Fig. \ref{fig:wII}, we can see that the dependence of frequencies of
$w_{\rm II}$ modes on the stellar compactness are almost independent from the values of ${\cal K}$
and the adopted EOS.
Additionally, it is found that the deviation from the frequencies expected in GR is very little.
We can see little difference between the expectations in GR and in TeVeS
in the damping rates (imaginary part of complex frequencies) for the stars
with larger compactness and in the oscillation frequencies (real part of complex frequencies) for the stars
with weaker compactness. For the oscillation frequencies in TeVeS,
we can get the empirical formula, such as
\begin{equation}
 Re\left(\omega M_{\rm ADM}\right) = -0.0412 + 0.7084 \left(\frac{M_{\rm ADM}}{R}\right)
     + 3.305 \left(\frac{M_{\rm ADM}}{R}\right)^2,
\end{equation}
and the oscillation frequencies can be in very good agreement with this empirical formula.
%
%
%

On the other hand, in Fig. \ref{fig:w1} for the frequencies of $w_1$ modes, we can observe same feature
as the frequencies of $w_{\rm II}$ modes, i.e., if we see those frequencies as functions of the stellar
compactness, those are almost independent from the values of ${\cal K}$ and the adopted EOS.
However, in the case of $w_1$ modes, there exists the crucial different feature in contrast to
the $w_{\rm II}$ modes. That is, the dependence of the frequencies in TeVeS is obviously different from those
expected in GR. In other words, with this different dependence of frequencies on the gravitational theory,
one can distinguish the gravitational theory in strong-field regime by using the gravitational waves
observations. In fact, the oscillation frequencies of axial $w_1$ modes in GR and TeVeS
can be expected with high accuracy via the following empirical formula
\begin{equation}
 Re\left(\omega R\right) = \alpha - \beta \left(\frac{M_{\rm ADM}}{R}\right),
\end{equation}
where $(\alpha,\beta)=(2.797,4.255)$ in GR and $(\alpha,\beta)=(2.533,3.714)$ in TeVeS.
%
%
%
%

\begin{figure}[htbp]
\begin{center}
\begin{tabular}{cc}
\includegraphics[scale=0.45]{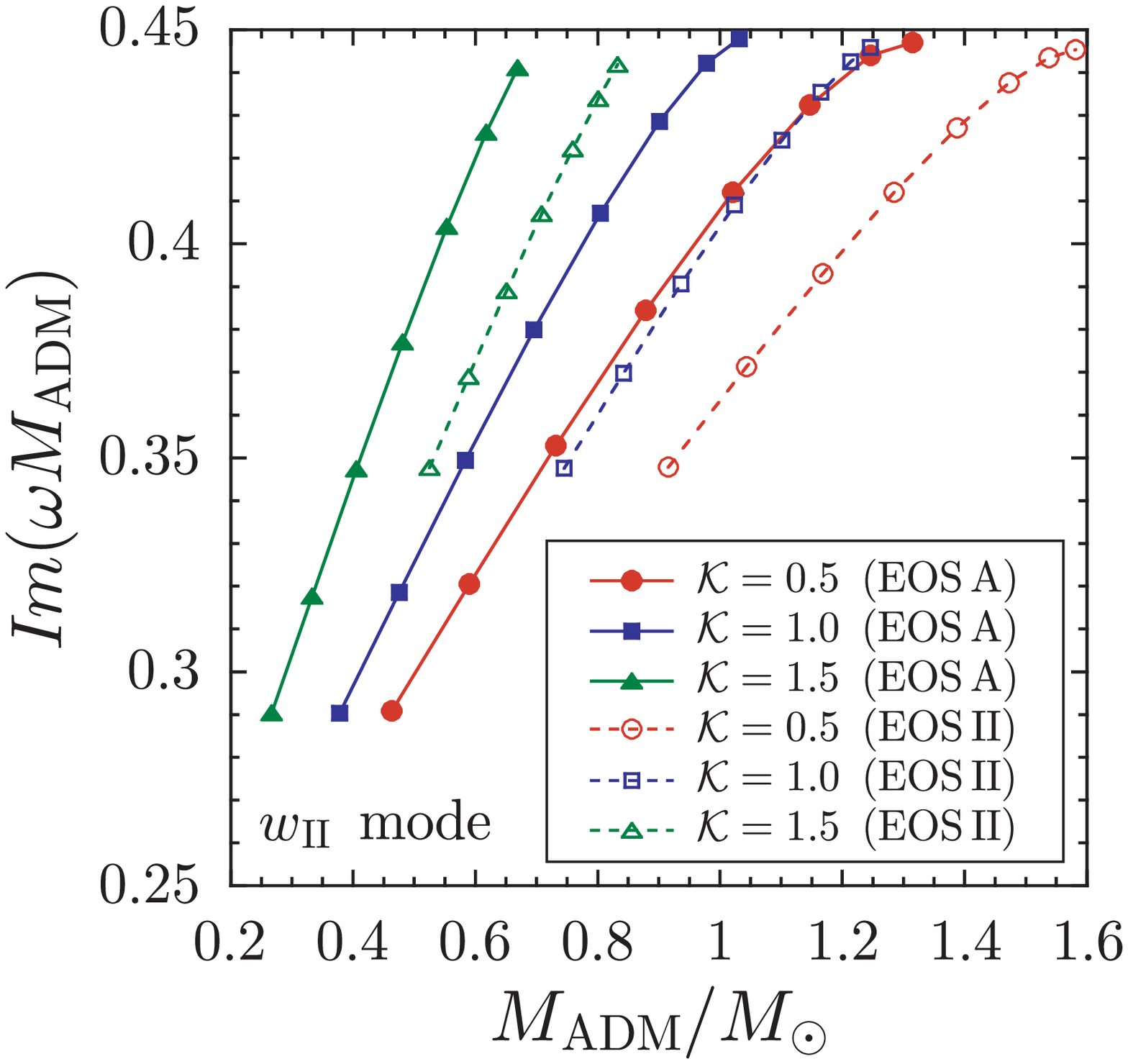} &
\includegraphics[scale=0.45]{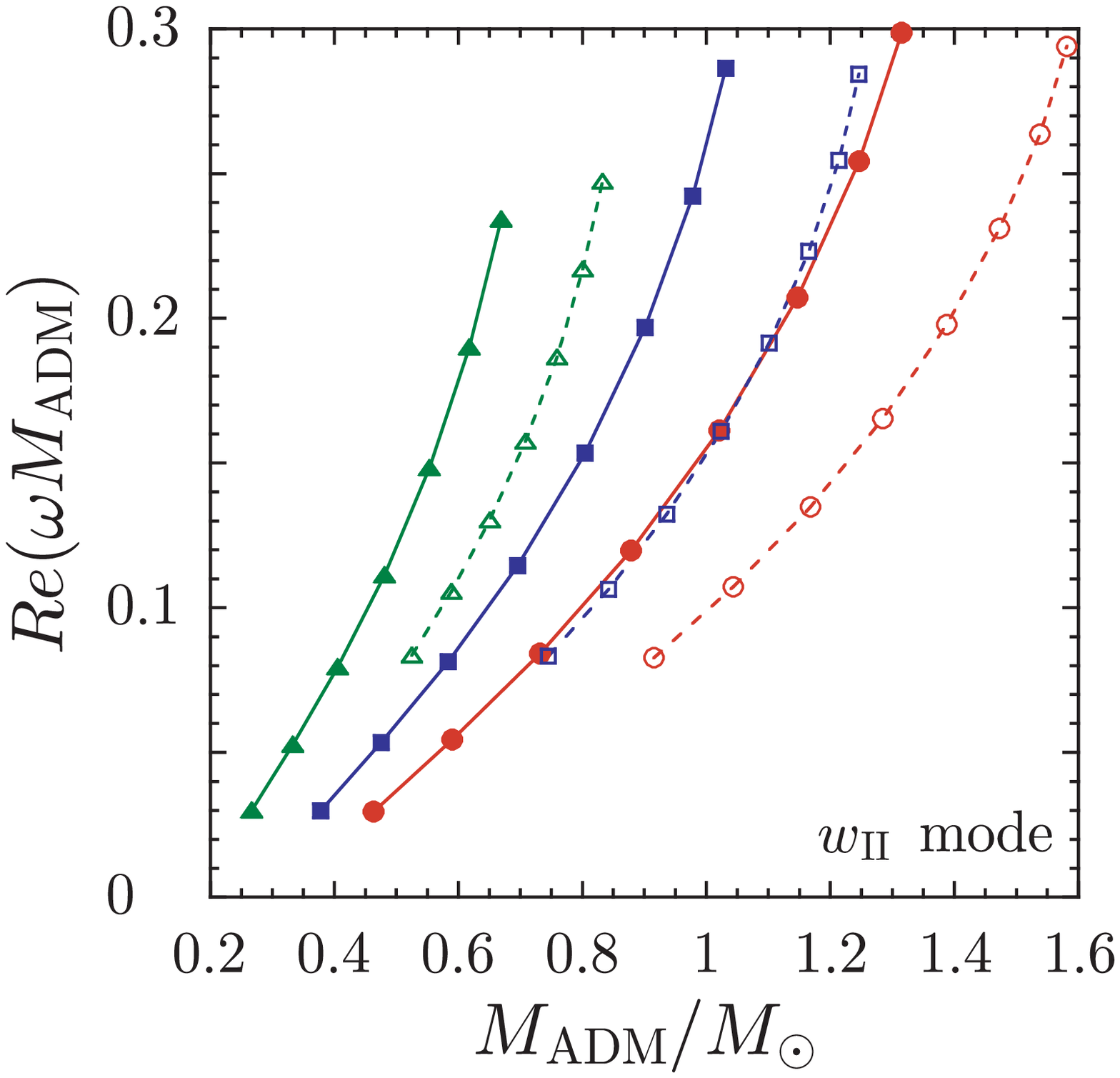} \\
\end{tabular}
\end{center}
\caption{
The normalized frequencies of $w_{\rm II}$ modes with two different EOSs are plotted as functions of
of the total ADM mass, where the left panel corresponds to the damping rates and the right panel are
to the oscillation frequencies. The solid lines denote the results with EOS A and the broken lines are
results with EOS II. Additionally, the different symbols denote the results with different values of
${\cal K}$, such as the circles for ${\cal K}=0.5$, the squares for ${\cal K}=1.0$, and the triangles
for ${\cal K}=1.5$.
}
\label{fig:wII-M}
\end{figure}

\begin{figure}[htbp]
\begin{center}
\begin{tabular}{cc}
\includegraphics[scale=0.45]{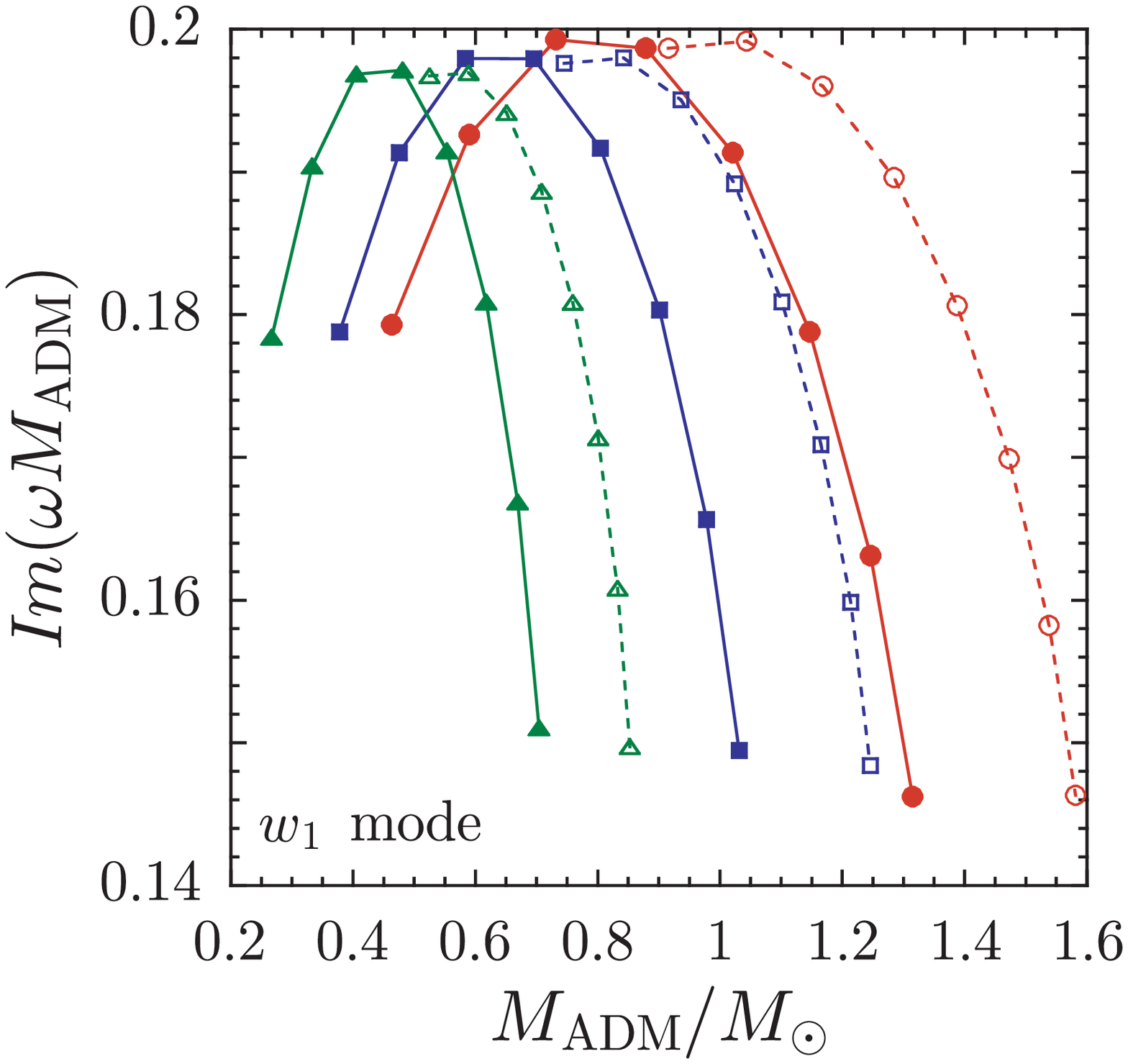} &
\includegraphics[scale=0.45]{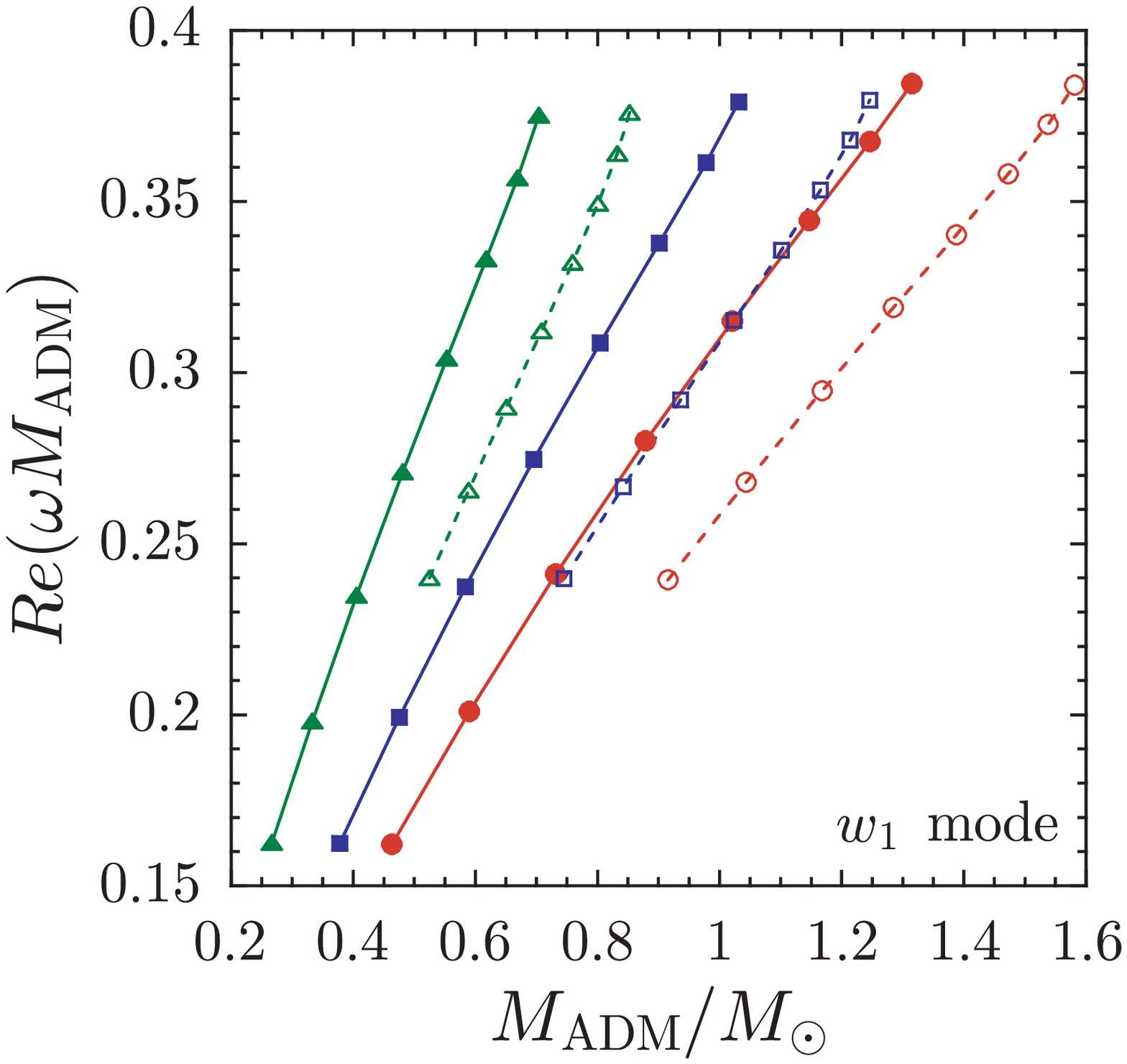} \\
\end{tabular}
\end{center}
\caption{
The normalized frequencies of $w_1$ modes with two different EOSs are plotted as functions of
of the total ADM mass, where the left panel corresponds to the damping rates and the right panel are
to the oscillation frequencies. The meaning of lines and symbols are same notations as Fig. \ref{fig:wII-M}.
}
\label{fig:w1-M}
\end{figure}

At the last, in order to discuss the possibility to make a restriction on the value of parameter ${\cal K}$,
we show the dependences of frequencies of $w_{\rm II}$ and $w_1$ modes on the total ADM mass in Figs.
\ref{fig:wII-M} and \ref{fig:w1-M}. These figures show
that due to the uncertainty of EOS it is not easy to determine the exact value of ${\cal K}$
by using the observations of axial $w_{\rm II}$ and $w_1$ modes.
However, with the help of the information of stellar mass,
combining the observations of axial gravitational waves with those of frequencies of fluid oscillations,
which depend strongly on the stellar average density \cite{Sotani2009a},
it might be possible to make a kind of constraint on the parameter ${\cal K}$ and/or on the stellar EOS.
%
%
%
%
%

\section{Conclusion}
\label{sec:V}

In order to examine how the tensor-vector-scalar (TeVeS) theory affects on the gravitational waves
emitted from the neutron stars, we derive the perturbation equations in TeVeS with the assumption
that the perturbations of vector field are neglected. Actually this assumption might not be physical, but with smaller value of the vector-field coupling constant ${\cal K}$ suggested by the observations of neutron star, the effect due to the presence of vector-field perturbation on the spacetime oscillation might be too small to neglect. Thus to see the dependence of spacetime oscillation on the gravitational theory, we adopt this simplification as a first step. With this assumption, we find that
the perturbations of scalar field couple with the polar perturbations of both spacetime and fluid,
but they do not affect on the axial perturbations. Since the axial spacetime modes are known to have
the same qualitative behavior as the polar spacetime modes
(at least in the case of GR), in this article we have studied only axial gravitational
waves, which are corresponding to the oscillations of spacetime itself.

With two different equations of state (EOS), we calculate the complex eigenfrequencies of axial
$w_{\rm II}$ and $w$ modes, where the real and imaginary parts of complex frequencies are corresponding
to the oscillation frequencies and the damping rates of the emitted gravitational waves, respectively.
We find that the dependences of the both frequencies of $w_{\rm II}$ and $w_1$ modes in TeVeS
on the stellar compactness are almost independent from the parameter ${\cal K}$ and the adopted EOS.
The dependences of $w_1$ modes are obviously different from that expected in GR,
while the dependences of $w_{\rm II}$ modes are similar to that in GR.
Owing to these differences of dependence on the gravitational theory,
one can distinguish the gravitational theory in the strong-field
regime by using the direct observations of gravitational waves emitted from the neutron stars.
Additionally, we show the possibility to make a kind of constraint on the value of ${\cal K}$ and/or
the stellar EOS via the observations of gravitational waves.

It should be noted that if one takes into account the effect of the vector-field perturbation, the quasinormal frequencies of spacetime oscillations could be changed with larger value of the coupling constant ${\cal K}$ although the qualitative behavior might be similar to our results in this paper. However, if our discussion would restrict in the region with smaller value or ${\cal K}$ as suggested in the neutron star observations, we could expect that the behavior of frequencies is almost same as that shown in this paper, i.e., we can see the obvious difference between the quasinormal frequencies expected in GR and in TeVeS. So still it could be possible to distinguish the gravitational theory with using the gravitational wave observations, although in this case it might be difficult to make a constraint on the value of ${\cal K}$.

In this article we focus only on the axial gravitational waves, but the study of polar gravitational waves
are also very interesting, since the gravitational waves couple directly with the perturbations of scalar field. 
Furthermore, since for simplicity we neglected the perturbations of vector field,
more detailed studies are needed including the perturbation of this field as well as those of metric and scalar fields.
Via these more detailed studies, we could obtain the additional information in the spectrum of emitted
gravitational waves
and one can provide more accurate constraints on the gravitational theory in the strong-field regime.
On the other hand, the consideration of the effects of the stellar magnetic field might be of great important. In fact,
recent observations of quasi-periodic oscillation in the giant flares of magnetar are believed to be related to the
oscillations of these strongly magnetized neutron stars \cite{Sotani2007,Sotani2008,Sotani2009b}.
By taking into account the effects of the stellar magnetic field, one might be able to set further constraints in the theory that describes gravity.

\acknowledgments

We thank K.D. Kokkotas for valuable comments.
This work was supported via the Transregio 7 ``Gravitational Wave Astronomy"
financed by the Deutsche Forschungsgemeinschaft DFG (German Research Foundation).

\appendix

\section{The Effects of Perturbation of Vector Field on the Spacetime Oscillations}   
\label{sec:appendix_0}

If one would consider the full problem where the perturbation of every field is taken into account,
the situation becomes more complicated. Although in this paper for simplicity we neglected the perturbation of the vector field such as $\delta {\cal U}_i=0$, in this appendix we want to see the effect of that perturbation on the metric perturbation briefly. In general, the perturbation of vector field could be given as
\begin{equation}
 \delta {\cal U}_\mu =\left(\frac{1}{2}e^{-\nu/2}\delta g_{tt},\delta{\cal U}_r,\delta{\cal U}_\theta,\delta {\cal U}_\phi\right),
\end{equation}
where the perturbed variables $\delta{\cal U}_r$, $\delta{\cal U}_\theta$, and $\delta {\cal U}_\phi$ are determined by calculating the linearized field equation for the vector field. With this expression of perturbation of vector field, the perturbed metric tensor in the Einstein frame is given by Eq.(\ref{p_Einstein}) as
\begin{align}
\delta g_{\mu\nu} = \delta g_{\mu\nu}^{(0)} +  (1-e^{4\varphi})e^{\nu/2}
 \sum_{l=2}^{\infty} \sum_{m=-l}^{l}\left(
 \begin{array}{cccc}
 0 & \delta {\cal U}_r
 & \delta {\cal U}_\theta
 & \delta {\cal U}_\phi \\
 \ast & 0 & 0 & 0 \\
 \ast & 0 & 0 & 0 \\
 \ast & 0 & 0 & 0 \\
 \end{array}
 \right),
\label{Eq:h1_Einstein}
\end{align}
where $\delta g_{\mu\nu}^{(0)}$ denotes the perturbed metric tensor in Einstein frame with the assumption that $\delta {\cal U}_i=0$, which is corresponding to Eq. (\ref{Eq:h_Einstein}). Thus if one considers the coupling of the perturbation of vector field with the perturbations of the other fields, the both axial and polar perturbations are connected to the vector-field perturbations. Namely, the axial spacetime modes should be also calculated together with the perturbation of vector field. However as mentioned in the main text, since the observations of neutron stars suggest that the value of the coupling constant for vector field, ${\cal K}$, should be smaller than at most ${\cal K}\simeq0.8$, the effect of the vector-field perturbation on the spacetime oscillations might be small. So as a first step we neglect the effect of the vector-field perturbation in this paper.

\section{The Components of the Linearized Einstein Equations}   
\label{sec:appendix_1}

In this appendix, we provide the explicit form of the various expressions used in Eqs.
(\ref{t-theta}) -- (\ref{phiphi})
for the linearized Einstein equations.
\begin{align}
 \alpha_{lm}^{(0)} =& \frac{1}{2}e^{-\zeta}\left[\hat{H}_1'-e^{\zeta}\left(\dot{\hat{H}}_2
     + \dot{\hat{K}}\right)+\frac{1}{2}\left(\nu' - \zeta'\right)\hat{H}_1\right]
     + 8\pi Ge^{-6\varphi}\left(\tilde{\rho}+\tilde{P}\right)V, \\
 \alpha_{lm}^{(1)} =& \frac{1}{2}\left[\hat{H}_0' - \hat{K}' - e^{-\nu}\dot{\hat{H}}_1
     + \left(\frac{1-{\cal K}}{2}\nu'-\frac{1}{r}\right)\hat{H}_0
     + \left(\frac{\nu'}{2}+\frac{1}{r}\right)\hat{H}_2 - \frac{16\pi}{k}\psi\delta\varphi\right], \\
 \beta_{lm}^{(0)} =& \frac{1}{2}e^{-\zeta}\left(\hat{h}_0''-\dot{\hat{h}}_1'\right)
     - \frac{1}{4}e^{-\zeta}\left(\nu'+\zeta'\right)\left(\hat{h}_0'-\dot{\hat{h}}_1\right)
     - \frac{1}{r}e^{-\zeta}\dot{\hat{h}}_1
     - 8\pi G e^{-6\varphi}\left(\tilde{\rho}+\tilde{P}\right)u \nonumber \\
    &- \bigg[\frac{l(l+1)}{2r^2} + \frac{1}{4}e^{-\zeta}\left({\nu'}^2+2\nu''\right)
     - \frac{2}{r^3}\left(m+rm'\right) + \frac{\nu'}{2r^2}\left(2r-3m-rm'\right)
       \nonumber \\
    &\hspace{1cm}
     + 8\pi Ge^{-6\varphi}\left(\tilde{\rho}+\tilde{P}\right) 
     - 16\pi G\tilde{P}e^{-2\varphi}\bigg]\hat{h}_0, \label{beta0} \\
 \beta_{lm}^{(1)} =& \frac{1}{2}e^{-\nu}\left(\dot{\hat{h}}_0'-\ddot{\hat{h}}_1\right)
     - \frac{1}{r}e^{-\nu}\dot{\hat{h}}_0 \nonumber \\
    &- e^{-\zeta}\left[\frac{(l-1)(l+2)}{2r^2}e^\zeta + \frac{3\nu'-\zeta'}{2r}
     - \frac{\nu'\zeta'}{4} + \frac{{\nu'}^2}{4} + \frac{\nu''}{2} - \frac{2m}{r^3}e^{\zeta}
     - 16\pi G\tilde{P}e^{-2\varphi+\zeta}\right]\hat{h}_1, \label{beta1} \\
 s_{lm} =& \frac{1}{2}\left(\hat{H}_0-\hat{H}_2\right), \\
 t_{lm} =& e^{-\nu}\dot{\hat{h}}_0-e^{-\zeta}\hat{h}_1'
     - \frac{1}{2}e^{-\zeta}\left(\nu'-\zeta'\right)\hat{h}_1, \label{tlm}
\end{align}
where $\psi\equiv \varphi'$.

\section{Numerical techniques}   
\label{sec:appendix_2}

Integrating from stellar center outward, one can get the values of $\Phi$ and $\Phi'$
at some radial position $r=r_a$ where is outside the star. With these values, we will have to match
the numerical solution with the appropriate asymptotic boundary conditions, which is the absence
of the incoming radiation. In order to find the asymptotic form of the solution of Eq. (\ref{eq:wave2})
with $\tilde{P}=0$, we can assume a solution of the form
\begin{equation}
 \Phi(r) = \left(\frac{r}{2m_c}-1\right)^{-2i\omega m_c}e^{-i\omega r}
     \sum_{n=0}^{\infty}a_n\left(1-\frac{r_a}{r}\right)^n,
\end{equation}
where $m_c$ is the mass function evaluated at radial infinity.
Substituting this form of the solution into the perturbation equation (\ref{eq:wave2}) and keeping
the terms up to the order $1/r^2$, we obtain a five-term recurrence relation for the expansion coefficients
$a_n$ for $n\ge 1$
\begin{equation}
 \alpha_n a_{n+1} + \beta_n a_n + \gamma_n a_{n-1} + \delta_n a_{n-2} + \epsilon_n a_{n-3} =0,
     \label{recurrence}
\end{equation}
where the coefficients of the recurrence relation are given by the following formulas
\begin{gather}
 \alpha_n   = c_0n(n+1), \\
 \beta_n    = d_0n + c_1n(n-1), \\
 \gamma_n   = e_0 + d_1(n-1) + c_2(n-1)(n-2), \\
 \delta_n   = e_1 + d_2(n-2) + c_3(n-2)(n-3), \\
 \epsilon_n = e_2 + d_3(n-3) + c_4(n-3)(n-4).
\end{gather}
In the above formulas, the coefficients $c_i$, $d_i$, and $e_i$ are functions of the
background quantities and have the form
\begin{gather}
 c_4 = -\frac{2m_1}{{r_a}^2}, \\
 c_3 = \frac{2m_c}{r_a} + \frac{8m_1}{{r_a}^2}, \\
 c_2 = 1-\frac{6m_c}{r_a}-\frac{12m_1}{{r_a}^2}, \\
 c_1 = -2 +\frac{6m_c}{r_a}+\frac{8m_1}{{r_a}^2}, \\
 c_0 = 1-\frac{2m_c}{r_a}-\frac{2m_1}{{r_a}^2}, \\
 d_3 = -\frac{6m_1}{{r_a}^2}\left(2i\omega m_c + 1\right), \\
 d_2 = \frac{4i\omega m_1}{r_a} + \frac{6m_c}{r_a} - 3d_3, \\
 d_1 = 2 - 2d_2 - 3d_3, \\
 d_0 = -2i\omega r_a-2 + d_2 + 2d_3, \\
 e_2 = \frac{24m_c^2m_1\omega^2}{{r_a}^2} + \frac{8im_c^3\omega}{{r_a}^2}
     + \frac{2}{{r_a}^2}\left(3m_1-m_c^2\right), \\
 e_1 = - \frac{8m_c\omega^2}{r_a}\left(3m_c^2 + m_1\right)
     + \frac{2im_1\omega}{r_a} - \frac{6m_c}{r_a} -2 e_2, \\
 e_0 = 2m_1\omega^2 - l(l+1) - e_1 - e_2,
\end{gather}
where $m_1$ is the coefficient in the asymptotic form of $m(r)$,
i.e., $m = m_c + m_1/r + {\cal O}(1/r^2)$,
which is given by
\begin{gather}
 m_1       = -\frac{kG^2M_\varphi^2}{8\pi} + \frac{{\cal K}}{4}m_c^2.
\end{gather}
%
%
The first four terms of the recurrence relation (\ref{recurrence}), i.e., $a_{-2}$, $a_{-1}$,
$a_0$, and $a_1$, are provided by the values of $\Phi$ and $\Phi'$ at $r=r_a$ as
\begin{gather}
 a_{-2} = a_{-1} = 0, \\
 a_0    = \frac{\Phi(r_a)}{X(r_a)}, \\
 a_1    = \frac{r_a}{X(r_a)}\left[\Phi'(r_a)+\frac{i\omega r_a}{r_a-2m_c}\Phi(r_a)\right],
\end{gather}
where
\begin{equation}
 X(r) = \left(\frac{r}{2m_c}-1\right)^{-2i\omega m_c}e^{-i\omega r}.
\end{equation}

In general, a high order recurrence relation can be reduced to a three-term recurrence relation,
in which case a convergence criteria for the solution can be applied, and one can identify
the solution describing only outgoing radiation \cite{Leaver1985}. In order to obtain a
three-term recurrence relation, we introduce new coefficients $\hat{\alpha}_n$, $\hat{\beta}_n$,
$\hat{\gamma}_n$, and $\hat{\delta}_n$ as
\begin{gather}
 \hat{\alpha}_1 = \alpha_1,\ \hat{\beta}_1 = \beta_1,\ \hat{\gamma}_1 = \gamma_1, \\
 \hat{\alpha}_2 = \alpha_2,\ \hat{\beta}_2 = \beta_2,\ \hat{\gamma}_2 = \gamma_2, \ \hat{\delta}_2 = \delta_2, 
\end{gather}
and for $n\ge 3$
\begin{gather}
 \hat{\alpha}_n = \alpha_n, \\
 \hat{\beta}_n  = \beta_n  - \frac{\hat{\alpha}_{n-1}\epsilon_n}{\hat{\delta}_{n-1}}, \\
 \hat{\gamma}_n = \gamma_n - \frac{\hat{\beta}_{n-1} \epsilon_n}{\hat{\delta}_{n-1}}, \\
 \hat{\delta}_n = \delta_n - \frac{\hat{\gamma}_{n-1}\epsilon_n}{\hat{\delta}_{n-1}}.
\end{gather}
Then the original five-term recurrence relation (\ref{recurrence}) becomes a four-term
recurrence relation for $n\ge 1$
\begin{equation}
 \hat{\alpha}_n a_{n+1} + \hat{\beta}_n a_n + \hat{\gamma}_n a_{n-1} + \hat{\delta}_n a_{n-2} = 0.
     \label{4-term}
\end{equation}
Furthermore, defining another set of coefficients $\tilde{\alpha}_n$, $\tilde{\beta}_n$, and
$\tilde{\gamma}_n$ as
\begin{equation}
 \tilde{\alpha}_1 = \hat{\alpha}_1,\ \tilde{\beta}_1 = \hat{\beta}_1,\ \tilde{\gamma}_1 = \hat{\gamma}_1,
\end{equation}
and for $n\ge 2$
\begin{gather}
 \tilde{\alpha}_n = \hat{\alpha}_n, \\
 \tilde{\beta}_n  = \hat{\beta}_n  - \frac{\tilde{\alpha}_{n-1}\hat{\delta}_n}{\tilde{\gamma}_{n-1}}, \\
 \tilde{\gamma}_n = \hat{\gamma}_n - \frac{\tilde{\beta}_{n-1} \hat{\delta}_n}{\tilde{\gamma}_{n-1}},
\end{gather}
the four-term recurrence relation (\ref{4-term}) can be reduced to a three-term recurrence relation
for $n\ge 1$
\begin{equation}
 \tilde{\alpha}_n a_{n+1} + \tilde{\beta}_n a_n + \tilde{\gamma}_n a_{n-1} =0.
\end{equation}
Using this three-term recurrence relation, the boundary condition at radial infinity
can be expressed as a continued fraction relation between $\tilde{\alpha}_n$, $\tilde{\beta}_n$,
and $\tilde{\gamma}_n$ as
\begin{equation}
 \frac{a_1}{a_0} = \frac{-\tilde{\gamma}_1}{\tilde{\beta}_1-}
    \frac{\tilde{\alpha}_1\tilde{\gamma}_2}{\tilde{\beta}_2-}
    \frac{\tilde{\alpha}_2\tilde{\gamma}_3}{\tilde{\beta}_3-}\cdots.
\end{equation}
Since this relation can be rewritten as 
\begin{equation}
 0 = \tilde{\beta}_0 - \frac{\tilde{\alpha}_0\tilde{\gamma}_1}{\tilde{\beta}_1-}
    \frac{\tilde{\alpha}_1\tilde{\gamma}_2}{\tilde{\beta}_2-}
    \frac{\tilde{\alpha}_2\tilde{\gamma}_3}{\tilde{\beta}_3-}\cdots \equiv f(\omega),
\end{equation}
where $\tilde{\beta}_0 \equiv a_1/a_0$ and $\tilde{\alpha}_0\equiv -1$,
hence, one can determine the eigenfrequnecy $\omega$ of a quasinormal mode by solving
the equation $f(\omega)=0$.


\end{document}